\documentclass{article}
\usepackage{spconf,amsmath,graphicx}

\usepackage{bbding, makecell}
\usepackage{hyperref}
\usepackage{amsfonts}
\usepackage{romannum}
\usepackage{makecell}
\usepackage{multirow}
\usepackage{xcolor}  
\usepackage{colortbl}  
\usepackage{threeparttable}
\newcommand{\tabincell}[2]{\begin{tabular}{@{}#1@{}}#2\end{tabular}}

\title{StuArt: Individualized Classroom Observation of Students with Automatic Behavior Recognition and Tracking}

\name{Huayi Zhou$^{\star}$ \quad Fei Jiang$^{\dagger}$ \sthanks{* Corresponding author: fjiang@mail.ecnu.edu.cn.} \quad Jiaxin Si$^{\S}$ \quad Lili Xiong$^{\ddag}$ \quad Hongtao Lu$^{\star}$ \thanks{This paper is supported by NSFC (No. 62176155, 62207014), Shanghai Municipal Science and Technology Major Project (2021SHZDZX0102). Hongtao Lu is also with MOE Key Lab of Artificial Intelligence, AI Institute, Shanghai Jiao Tong University.}}
\address{$^{\star}$ Department of Computer Science and Engineering, Shanghai Jiao Tong University; \\
    $^{\dagger}$ Shanghai Institute of AI Education, East China Normal University; \\
    $^{\S}$ Chongqing Qulian Digital Technology; $^{\ddag}$ Chongqing Academy of Science and Technology}

\begin{document}
%
\maketitle
%
\begin{abstract}

Each student matters, but it is hardly for instructors to observe all the students during the courses and provide helps to the needed ones immediately. In this paper, we present StuArt, a novel automatic system designed for the individualized classroom observation, which empowers instructors to concern the learning status of each student. StuArt can recognize five representative student behaviors (hand-raising, standing, sleeping, yawning, and smiling) that are highly related to the engagement and track their variation trends during the course. To protect the privacy of students, all the variation trends are indexed by the seat numbers without any personal identification information. Furthermore, StuArt adopts various user-friendly visualization designs to help instructors quickly understand the individual and whole learning status. Experimental results on real classroom videos have demonstrated the superiority and robustness of the embedded algorithms. We expect our system promoting the development of large-scale individualized guidance of students. More information is in \url{https://github.com/hnuzhy/StuArt}.
\end{abstract}
\begin{keywords}
Behavior recognition, behavior tracking, classroom observation, education, individualization.
\end{keywords}
%
\section{Introduction}

Classroom observation serves as a means of assessing the effectiveness of teaching, which has been continuously studied \cite{adelman2003guide, wragg2011introduction, o2020classroom}. The utility of traditional classroom observation is limited due to the manual coding that is expensive and unscalable. Thus, automatic classroom observation systems built on the advanced technologies are quite essential.

Existing research of automatic classroom observation focuses on extracting features associated with effective instruction based on non-invasive audiovisual records or sensors, such as Sensi \cite{saquib2018sensei}, EduSense \cite{ahuja2019edusense} and ACORN \cite{ramakrishnan2021toward}. As shown in Table \ref{systems}, however, the above-mentioned systems are not completed. First, the isolated evaluation dimension reduces their application values. It is hard for teachers to rapidly obtain overall class tendency by only providing emotions or single behavior (e.g., affective \cite{littlewort2011computer, zeng2020emotioncues}, gaze \cite{ahuja2021classroom}, cognitive \cite{picard2011measuring, hassib2017engagemeter} and behavioral \cite{lin2018hand, ahuja2019edusense, zheng2020intelligent}). Second, individualized classroom observations that record the learning status of each student in class are rarely presented. It is not only an important indicator of teaching quality, but also empowers instructors to know each student for providing personalized instruction.

\setlength{\tabcolsep}{1pt}
\begin{table}[]\scriptsize
\caption{Comparison between our system and representative classroom observation literature related to students.}
\centering
\begin{threeparttable}[b]
\begin{tabular}{c|c|c|c|c|c}
	\Xhline{1.2pt}
	{\bf Literature} & {\bf Sensor} & \tabincell{c}{\bf Visual \\ \bf Tracking} & \tabincell{c}{\bf Individual\\ \bf Observation?} & \tabincell{c}{\bf Holistic\\ \bf Observation?} & \tabincell{c}{\bf Related \\ \bf Features\tnote{[a]}} \\
	\Xhline{1.2pt}
	Affectiva \cite{picard2011measuring} & Wrist-worn & \XSolidBrush & \Checkmark & \XSolidBrush & (5,10,11) \\
	CERT \cite{littlewort2011computer} & Ready videos & \Checkmark & \XSolidBrush & \Checkmark & (5,7,10)  \\
	EngageMeter \cite{hassib2017engagemeter} & Headsets & \XSolidBrush & \Checkmark & \XSolidBrush & (12) \\
	Sensei \cite{saquib2018sensei} & A radio net & \XSolidBrush & \XSolidBrush & \Checkmark & (8) \\
	EduSense \cite{ahuja2019edusense} & 2 RGB cams & \Checkmark & \XSolidBrush & \Checkmark & (1,2,5,7,8,9) \\
	EmotionCues \cite{zeng2020emotioncues} & Ready videos & \XSolidBrush & \XSolidBrush & \Checkmark & (10) \\
	Zheng et al. \cite{zheng2020intelligent}& 1 RGB cam & \XSolidBrush & \XSolidBrush & \Checkmark & (1,2,3) \\
	Ahuja et al. \cite{ahuja2021classroom} & 2 RGB cams & \Checkmark & \XSolidBrush & \Checkmark & (6) \\
	ACORN \cite{ramakrishnan2021toward} & Multimodel &\XSolidBrush & \XSolidBrush & \Checkmark & (13) \\
	\hline
	StuArt & 1 RGB cam & \Checkmark & \Checkmark & \Checkmark & (1,2,3,4,5,8)  \\
	\Xhline{1.2pt}
\end{tabular}
\begin{tablenotes}
\item [{[a]}] (1) hand-raising, (2) standing, (3) sleeping, (4) yawning, (5) smiling, (6) eye gaze, (7) head pose, (8) position, (9) speech, (10) emotion, (11) electrodermal activity, (12) electroencephalography (EEG), (13) overall classroom climate
\end{tablenotes} 
\end{threeparttable}
\label{systems}
\vspace{-20pt}
\end{table}

\begin{figure*}[!t]
\centering
\includegraphics[width=\textwidth]{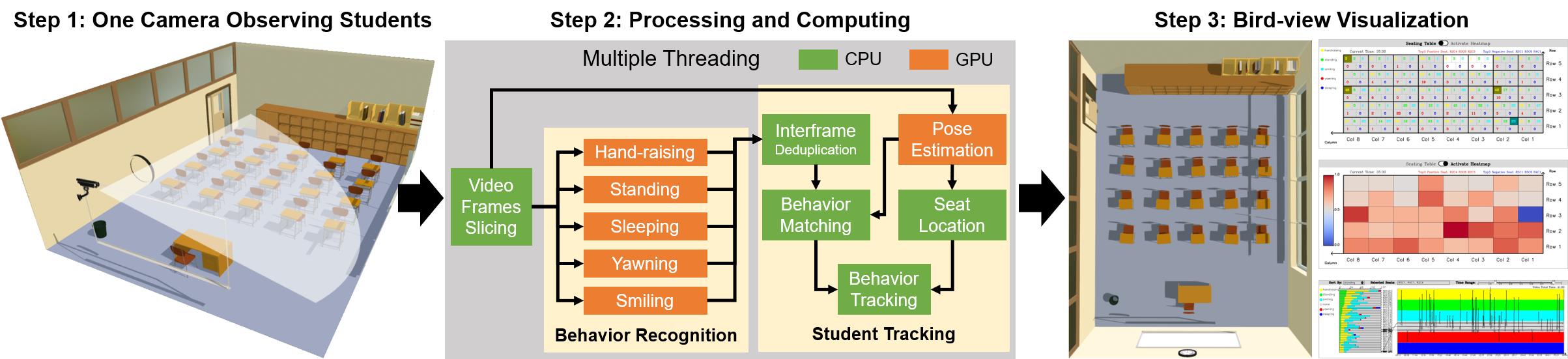}
\caption{Overall architecture of StuArt system. Detailed descriptions of the main components for each step are presented later.}
\label{stuart}
\vspace{-10pt}
\end{figure*}

In this paper, we present an automatic individualized classroom observation system StuArt using one RGB camera. First, to reflect the learning status of students, five typical behaviors that occur in class and their variation trends are presented, including hand-raising, standing, smiling, yawning and sleeping. To improve the behavior recognition performance, we build a large-scale student behavior dataset. Second, to observe students continuously, we propose a multiple-student tracking method. It utilizes seat locations (row and column) of students as their unique identifications to protect personal privacy. Moreover, we design several intuitive visualizations to display the evolution process of the individualized and holistic learning status of students.

As far as we know, StuArt is the first comprehensive system of automatic individualized classroom observation. All techniques embedded in it are designed for classroom scenarios, and systematically evaluated effects in both controlled and real courses. The robust performances prove the usability and generalization value of our system.

\section{System Design of StuArt}

Our proposed individualized classroom observation system, StuArt, is shown in Figure \ref{stuart}, which comprises three main steps. First, the data of students' learning process is recorded by one high-definition RGB camera. Then, the video is parsed to generate educationally-relevant features, and the variation of learning status for each student. Finally, we visualize the individual and holistic situations of students from a bird-view according to their seat positions.

{\bf \Romannum{1}. Equipment Overview:}
Equipment of StuArt has three aspects: video data capturing by the RGB camera, models training and system deployment in running device. We use one single camera to capture videos for two motivations. Firstly, CNN-based algorithms can detect objects quite accurately in real classrooms \cite{lin2018hand, ahuja2019edusense, zheng2020intelligent, ahuja2021classroom}. Secondly, camera deployment is portable, especially in the we-media era. A camera with 1K (1920$\times$1080) resolution costing about \$300 meets our requirements. For large-scale data processing and training, we equipped storage arrays and multiple GPUs. In inference, StuArt only needs one 3080Ti GPU. Considering computational cost, we also implemented an integrated system on the more lightweight edge device Jetson TX2 8GB module costing about \$500. 

\begin{figure}[!t]
\centering
\includegraphics[width=\columnwidth]{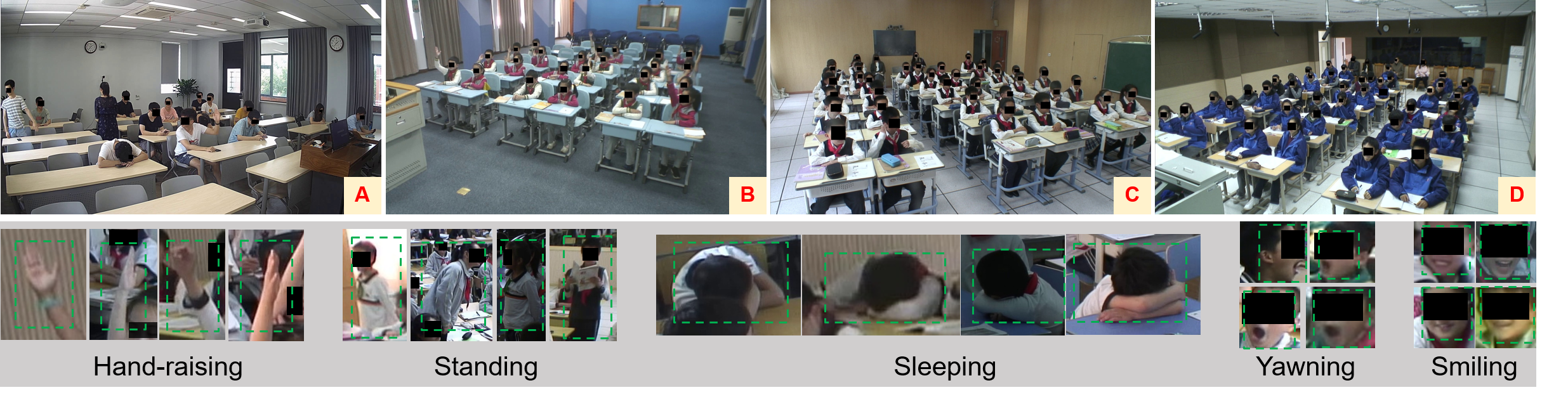}
\caption{{\bf Top:} examples of K-12 classroom scenarios. {\bf Bottom:} some samples of annotated behaviors.}
\label{classrooms}
\vspace{-10pt}
\end{figure}

{\bf \Romannum{2}. Real Instructional Videos:} 
We obtained the access rights of generous real instructional K-12 course videos, which are collected and managed by the institute of education in Shanghai city. Generally, we divide these courses into three categories including primary, junior and senior, shown by sub-images in Figure \ref{classrooms} B, C and D. In Figure \ref{classrooms} A, we also built our simulated teaching scene. As can be seen, students in classrooms are visually crowded, which brings low resolution and severe occlusion challenges in rear rows. Thus, the performance of deep vision algorithms trained on public datasets is often degraded. Domain behavior datasets are essential for adapting to the classroom scene.

Specifically, we selected about 300 videos from 20+ schools as the raw data. Each course video has an average duration of 40 minutes with 25fps. Firstly, we use OpenCV \cite{bradski2008learning} and FFMPEG \cite{tomar2006converting} to do preprocessing. Then, we set an interval at 3 seconds to evenly sample frames. We select five behaviors to annotate: {\bf hand-raising, standing, sleeping, yawning} and {\bf smiling}. To prevent the wrong recognition of standing with teachers walking into the student group, we add one more category namely {\bf teacher}. During annotation, we choose the tool LabelMe \cite{russell2008labelme} and store the location $(x, y, w, h)$ and category of behaviors in an XML file as Pascal VOC \cite{everingham2010pascal}. In order to ensure the labeling accuracy and efficiency, we finished annotation by our team's researchers familiar with the goal following the active-learning strategy. Some annotation examples are in Figure \ref{classrooms} bottom.

\begin{figure}[]
\centering
\includegraphics[width=\columnwidth]{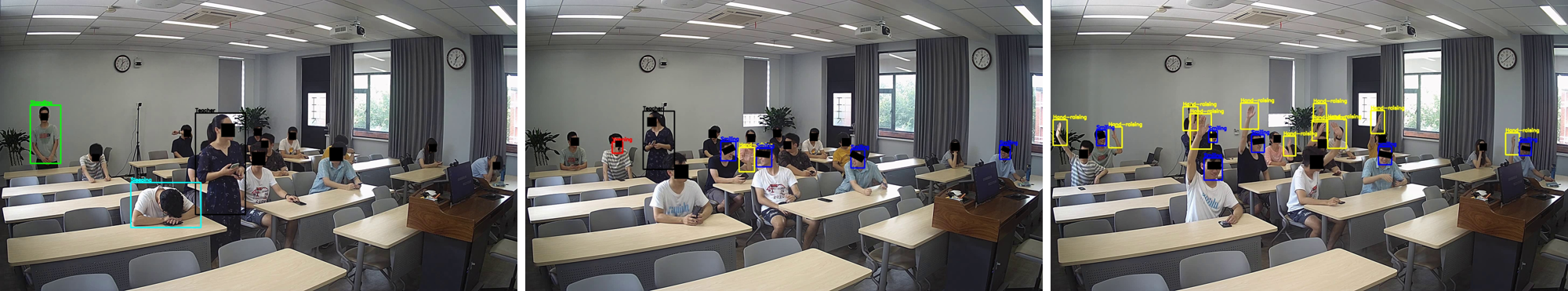}
\caption{Qualitative behavior detection results. Yellow, green, cyan, red, blue and black boxes are hand-raisings, standing, sleeping, yawning, smiling and teacher, respectively.}
\label{behaviorDet}
\vspace{-10pt}
\end{figure}
{\bf \Romannum{3}. Behavior Recognition:} 
We treat the behavior recognition task as the object detection problem using YOLOv5 \cite{jocher2020yolov5}, which keeps a better trade-off between accuracy and computation than its counterparts \cite{ren2015faster, dai2016r}. Figure \ref{behaviorDet} shows detection examples. Due to the size gap between yawning and smiling with other three behaviors, we failed to incorporate them into the multi-behavior detector. For yawning, we allocated a single detector and proposed a mouth fitting strategy to distinguish it with talking. For smiling, we adopted face detection MTCNN \cite{zhang2016joint} and emotion classification CovPoolFER \cite{acharya2018covariance} to detect it. By simply reducing YOLOv5 from large to small, we obtain light-weight student behavior detectors for running in edge devices.

{\bf \Romannum{4}. Student Tracking:}
This part, we present how to reasonably count the times of behaviors, match behaviors with individual students, and then track the behavior sequence of the whole class according to the seat of each student.

{\it \romannum{1}. Interframe Deduplication:} The input of behavior detector is sampled frame without temporal information. However, a real behavior may last across consecutive frames. Thus, we need to suppress duplications. We adopt bounding box matching to filter out repeated behaviors with maximum IoU among interframes. In practice, we set our overlap threshold to $0.2$ for robust IoU matching. If a behavior box is not matched in subsequent $T$ frames, we assume it happened once. We here set $T=2$ in case of occasional missed behavior detection.

\begin{figure}[!t]
\centering
\includegraphics[width=\columnwidth]{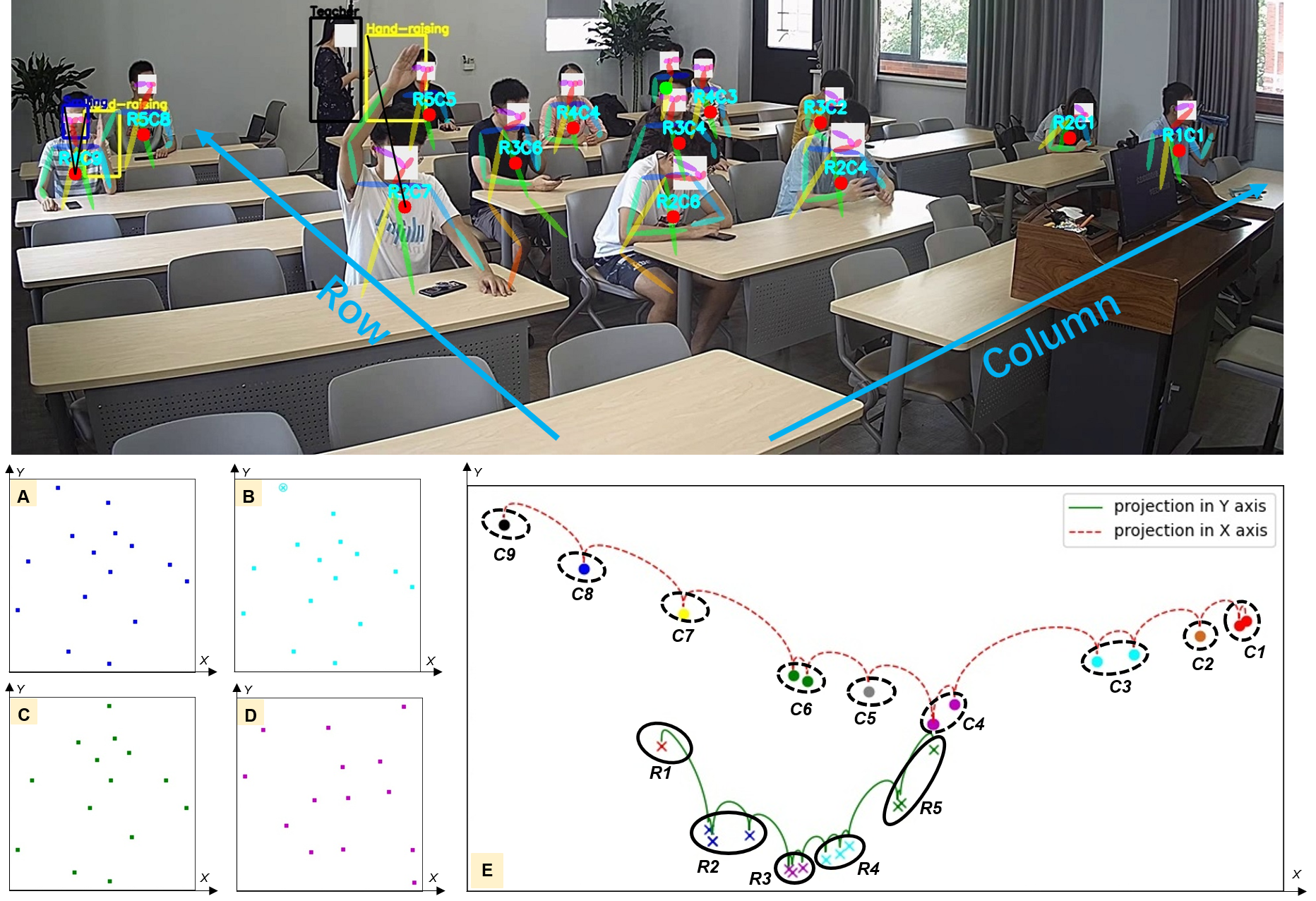}
\caption{{\bf Top:} colorful skeletons are detected by OpenPifPaf. Red solid circles are representative points of students. Black lines indicate behavior matching results. Cyan texts with format {\it RxCy} are row/column numbers of students used in seat location. {\bf Bottom:} illustrations of five steps in our seat location algorithm. Points are students in the top image.}
\label{behaviorLoc}
\vspace{-10pt}
\end{figure}

{\it \romannum{2}. Behavior Matching:} How do we know who are performing these behaviors? We need to complete the matching of behaviors. Except for hand-raising, the other four behaviors have naturally identified the corresponding students with their bounding boxes (refer Figure \ref{classrooms}). The hand-raising box only contains partial area of the raising arm, which is likely to be confused with the wrong individual student's body when matching, especially the simultaneous occurrence of dense hand-raisings among crowds \cite{zhou2018raising, narasimhaswamy2022whose}. To tackle it, we adopted OpenPifPaf \cite{kreiss2021openpifpaf} to generate student body keypoints as auxiliary information for hand-raising matching. We plotted skeletons for illustrating in Figure \ref{behaviorLoc} top. However, due to inherent challenges of hand-raising gestures, OpenPifPaf may still fail to detect arm joints. Thus, we designed a heuristic strategy combining greedy pose searching and weight allocation priority to maximize the matching accuracy \cite{zhou2018raising}.

{\it \romannum{3}. Seat Location:} After finishing behavior matching, we still keep unknown about who or where they are. We elegantly chose to use the row and column number of student seats to uniquely represent them. Concretely, we divide the seat location into five steps shown in Figure \ref{behaviorLoc} bottom: (A) Based on body keypoints, we generate representative points by averaging coordinates of upper body joints. (B) With these points, we filter out the teacher point if detected. (C) Then, if using a wide-angle camera, we need to correct the barrel distortion of these points. (D) After that, to make 2D points under camera view with affine deformation in distinct order, we rearrange them from top-view by applying projective transformation \cite{bradski2008learning}. (E) Finally, the row and column can be obtained by applying K-means clustering after projecting points on the $y$-axis and $x$-axis, respectively. In Figure \ref{behaviorLoc}, all students' located seats are plotted. We can automatically recognize that two students sitting in {\it R2C7} and {\it R4C9} are raising his or her hand. And a smiling student is sitting in {\it R4C9}.

{\it \romannum{4}. Behavior Tracking:} By now, we can automatically recognize student behaviors, suppress repeated behaviors, match behaviors with students, and locate student seats with given total rows and columns. Based on these results, we could obtain individual behavior tracklets of every student. Their seating positions indexed by {\it RxCy} are regarded as  tracklet IDs. In this way, if a classroom equipped with $i$ rows and $j$ columns seats, we would totally generate $i \times j \times 5$ tracklets for tracking five specific behaviors. 

{\bf \Romannum{5}. Individualized Observation Visualization:} 
To assist teachers quickly observe each student's status and know the classroom situation, we designed visualizations from both micro (individual student) and macro (overall students) levels. 

\begin{figure}
\centering
\includegraphics[width=\columnwidth]{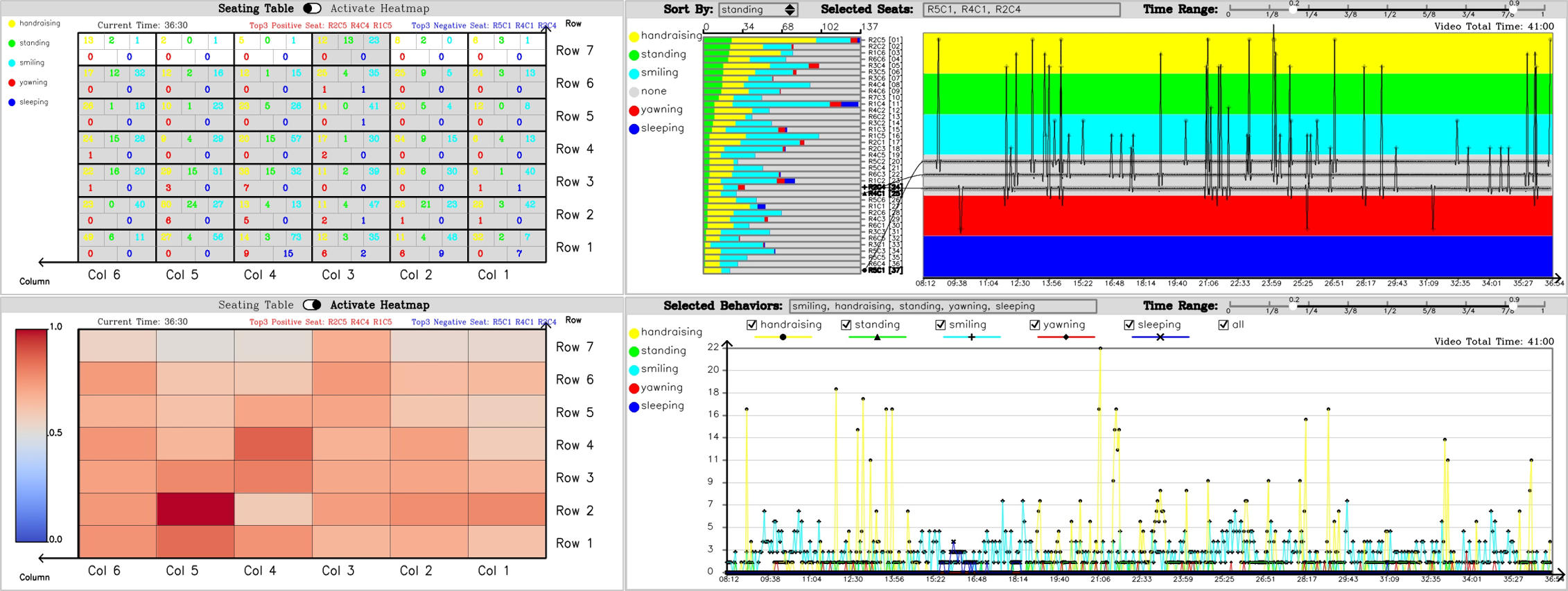}
\caption{{\bf Left:} visualizations of two non-interactive designs. The current data corresponds to the frame shown in Figure \ref{behaviorLoc} top. {\bf Right:} visualizations of two non-interactive designs.}
\label{visualization}
\vspace{-10pt}
\end{figure}

{\it \romannum{1}. Non-interactive Designs:}
As shown in Figure \ref{visualization} top-left ({\bf Seating Table}), using our simulated video as an example, we draw and update the number of five behaviors in all table cells representing student seats as time goes on. The seats occupied by students have gray background. Behavior categories are distinguished by different font colors. If a behavior is detected, the corresponding table cell will be highlighted with background color related to its category. Furthermore, as in Figure \ref{visualization} bottom-left ({\bf Activate Heatmap}), by assuming that the counts of positive (hand-raising, standing and smiling) and negative (yawning and sleeping) behaviors roughly reflects engagement, we can get the dynamic tendency of students' engagement by converting behavior counts into the range $[0, 1]$ for rendering a heatmap with diverse cold and warm colors ranging between blue and red. These presented details are synchronized with class time, and the detection and tracking results will be refreshed automatically every three seconds. No intervention required makes this way suitable for being used when in class.

{\it \romannum{2}. Interactive Designs:}
After class, the non-interactive designs may be unsatisfactory for reviewing a student in detail. Inspired by EmotionCues \cite{zeng2020emotioncues}, we designed configurable interactive {\bf Link List} and {\bf Point Flow} based on the recorded behavior sequence of each student, as shown in Figure \ref{visualization} right. In each link list, we arrange student behavior counts into a list of horizontally stacked bars on the left space. Teachers can re-sort them according to the target behavior type. These bars are clickable and linked with detailed personal behavior sequences. Instead of observing an isolated student, we can also present the integrated information of the classroom by depicting the whole behavior trend at each sampling moment forming the point flow. This design is a supplement to the previous link list interaction.

\section{Experiments}

\subsection{Data for Training and Testing}

We have obtained a total of nearly 36k images which include more than 110k instances. Specifically, for the training of multi-objects model including hand-raising, standing, sleeping and teacher, we have gotten about 70k, 21k, 3k and 15k samples respectively. The number of images and samples used for training single yawning detection model are 3,162 and 3,216. For smiling, we got a total of 129k positive and 181k negative facial images. Additionally, to test the accuracy of the two vital parts (behavior matching and seat location) in student tracking, we selected six and four courses respectively. Among them, the six courses used to evaluate matching include primary, junior and senior schools, as well as left and right observation perspectives. The four videos used to test the seat location effect include distorted and undistorted videos, as well as left and right views.

\subsection{Effectiveness Evaluation}

{\bf Behavior Recognition:} We adopt the metric mAP (mean Average Precision) which is widely used in public object detection datasets Pascal VOC \cite{everingham2010pascal} and COCO \cite{lin2014microsoft} to evaluate our behavior recognition methods. On the behavior validation set, our multi-objects model which contains the detection of hand-raising, standing, sleeping and teacher achieved {\bf 57.6} mAP(\%). The model used to detect yawning alone reaches {\bf 90.0} mAP(\%). For smiling recognition, we got an average of {\bf 84.4\%} accuracy for binary classification.

{\bf Behavior Matching:} We only report the evaluation of the most difficult hand-raising behavior matching on six selected courses. We automatically detected and obtained 2,667 hand-raising bounding boxes in total, and corresponding matched box number is 2,625. Among them, we manually figured out 2,409 real hand-raising boxes with 2,001 correctly matched. The final behavior matching precision and recall are {\bf 83.06\%} (2001/2409) and {\bf 98.43\%} (2625/2667) respectively, which far exceeds the results by applying unoptimized matching strategy (which has 68.04\% precision and 83.39\% recall).

{\bf Seat Location:} Table \ref{tab2} shows quantitative details of the seat location on four selected courses. We randomly chose five seats from $RxCy$ in each class, and observed the location result of each. The higher the values of $x$ and $y$, the farther away from the camera the students are, and the lower their resolution is. Then, we manually counted out $F_{l}$ legal frames in which student's location is distinguishable, and  $F_{c}$ frames where students are correctly located. The location accuracy $Acc_{s}$ of a single student in one course can be roughly calculated by $F_{c}/F_{l}$. Finally, the overall student location accuracy $Acc_{a}$ is about {\bf 83.3\%}. Although the location accuracy of some students (e.g., $R4C3$ in video 2, $R5C5$ in video 3) is limited, most of them are located well enough to support rough student tracking and individualized observation.

\setlength{\tabcolsep}{0.7pt}
\begin{table}[]\scriptsize
\begin{center}
\caption{Seat location results on four selected course videos.}
\label{tab2}
\begin{tabular}{c|c|c|c|c|c|c|c|c|c|c|c|c|c|c|c|c|c|c|c|c|c}
\Xhline{1.5pt}
{\bf \makecell{Video \\ ID}} & \multicolumn{5}{c|}{\bf \makecell{1 \\ undistorted \\ left view}} & \multicolumn{5}{c|}{\bf \makecell{2 \\ undistorted \\ right view}} & \multicolumn{5}{c|}{\bf \makecell{3 \\ undistorted \\ left view}} & \multicolumn{5}{c|}{\bf \makecell{4 \\ distorted \\ left view}} & {\bf Total} \\
\Xhline{1.2pt}
\makecell{Total \\ R/C} & \multicolumn{5}{c|}{\makecell{6 Rows \\ 6 Columns}} & \multicolumn{5}{c|}{\makecell{5 Rows \\ 8 Columns}} & \multicolumn{5}{c|}{\makecell{5 Rows \\ 7 Columns}} & \multicolumn{5}{c|}{\makecell{5 Rows \\ 6 Columns}} & \\
\hline
{\it RxCy} & \tabincell{c}{R1\\C2} & \tabincell{c}{R2\\C1} & \tabincell{c}{R3\\C4} & \tabincell{c}{R5\\C5} & \tabincell{c}{R6\\C5} & \tabincell{c}{R2\\C2} & \tabincell{c}{R1\\C5} & \tabincell{c}{R4\\C3} & \tabincell{c}{R4\\C6} & \tabincell{c}{R3\\C8} & \tabincell{c}{R1\\C1} & \tabincell{c}{R2\\C3} & \tabincell{c}{R4\\C2} & \tabincell{c}{R3\\C6} & \tabincell{c}{R5\\C5} & \tabincell{c}{R1\\C2} & \tabincell{c}{R3\\C1} & \tabincell{c}{R3\\C3} & \tabincell{c}{R2\\C2} & \tabincell{c}{R5\\C6} & ~ \\
\hline
{${F}_{l}$} & \rotatebox[origin=c]{90}{719} & \rotatebox[origin=c]{90}{723} & \rotatebox[origin=c]{90}{704} & \rotatebox[origin=c]{90}{700} & \rotatebox[origin=c]{90}{710} & \rotatebox[origin=c]{90}{457} & \rotatebox[origin=c]{90}{459} & \rotatebox[origin=c]{90}{458} & \rotatebox[origin=c]{90}{436} & \rotatebox[origin=c]{90}{410} & \rotatebox[origin=c]{90}{839} & \rotatebox[origin=c]{90}{840} & \rotatebox[origin=c]{90}{868} & \rotatebox[origin=c]{90}{858} & \rotatebox[origin=c]{90}{742} & \rotatebox[origin=c]{90}{694} & \rotatebox[origin=c]{90}{652} & \rotatebox[origin=c]{90}{698} & \rotatebox[origin=c]{90}{694} & \rotatebox[origin=c]{90}{695} & \rotatebox[origin=c]{90}{{\bf 11123}} \\
\hline
{${F}_{c}$} & \rotatebox[origin=c]{90}{692} & \rotatebox[origin=c]{90}{702} & \rotatebox[origin=c]{90}{684} & \rotatebox[origin=c]{90}{662} & \rotatebox[origin=c]{90}{641} & \rotatebox[origin=c]{90}{350} & \rotatebox[origin=c]{90}{417} & \rotatebox[origin=c]{90}{295} & \rotatebox[origin=c]{90}{285} & \rotatebox[origin=c]{90}{323} & \rotatebox[origin=c]{90}{760} & \rotatebox[origin=c]{90}{704} & \rotatebox[origin=c]{90}{810} & \rotatebox[origin=c]{90}{628} & \rotatebox[origin=c]{90}{484} & \rotatebox[origin=c]{90}{603} & \rotatebox[origin=c]{90}{485} & \rotatebox[origin=c]{90}{542} & \rotatebox[origin=c]{90}{588} & \rotatebox[origin=c]{90}{468} & \rotatebox[origin=c]{90}{{\bf 13356}} \\
\hline
{${Acc}_{s}$} & \rotatebox[origin=c]{90}{0.96} & \rotatebox[origin=c]{90}{0.97} & \rotatebox[origin=c]{90}{\textcolor{red}{0.97}} & \rotatebox[origin=c]{90}{0.95} & \rotatebox[origin=c]{90}{\textcolor{blue}{0.90}} & \rotatebox[origin=c]{90}{0.77} & \rotatebox[origin=c]{90}{\textcolor{red}{0.91}} & \rotatebox[origin=c]{90}{\textcolor{blue}{0.64}} & \rotatebox[origin=c]{90}{0.65} & \rotatebox[origin=c]{90}{0.79} & \rotatebox[origin=c]{90}{0.91} & \rotatebox[origin=c]{90}{0.84} & \rotatebox[origin=c]{90}{\textcolor{red}{0.93}} & \rotatebox[origin=c]{90}{0.73} & \rotatebox[origin=c]{90}{\textcolor{blue}{0.65}} & \rotatebox[origin=c]{90}{\textcolor{red}{0.87}} & \rotatebox[origin=c]{90}{0.74} & \rotatebox[origin=c]{90}{0.78} & \rotatebox[origin=c]{90}{0.85} & \rotatebox[origin=c]{90}{\textcolor{blue}{0.67}} & {\bf --} \\
\hline
{${Acc}_{a}$} & \multicolumn{5}{c|}{\bf 95.1\%} & \multicolumn{5}{c|}{\bf 75.2\%} & \multicolumn{5}{c|}{\bf 81.2\%} & \multicolumn{5}{c|}{\bf 78.2\%} & {\bf 83.3\%}\\
\Xhline{1.5pt}
\end{tabular}
\end{center}
\vspace{-20pt}
\end{table}

\section{Conclusion}

In the traditional classroom for K-12 education, we proposed StuArt to realize individualized classroom observation by automatically recognizing and tracking student behaviors using one front-view camera. Firstly, we selected five typical classroom behaviors as entry perspectives. Then, we built a dedicated behavior dataset to accurately detect target behaviors. For behavior matching and seat location, we chose the classroom with a traditional seating arrangement, and used the row and column of the seat as a desensitized identification to realize the automatic tracking of student behavior. Based on these observation results, we have designed suitable visualizations to assist teachers quickly obtain student behavior patterns and the overall class state. In experiments, we conducted quantitative evaluations of the proposed core methods. The results confirmed the effectiveness of StuArt.

\bibliographystyle{IEEEbib}
\bibliography{refs}

\begin{thebibliography}{10}

\bibitem{adelman2003guide}
Clement Adelman, Clem Adelman, and Roy Walker,
\newblock {\em A guide to classroom observation},
\newblock Routledge, 2003.

\bibitem{wragg2011introduction}
Ted Wragg,
\newblock {\em An introduction to classroom observation (Classic edition)},
\newblock Routledge, 2011.

\bibitem{o2020classroom}
Matt O’Leary,
\newblock {\em Classroom observation: A guide to the effective observation of
  teaching and learning},
\newblock Routledge, 2020.

\bibitem{saquib2018sensei}
Nazmus Saquib, Ayesha Bose, Dwyane George, and Sepandar Kamvar,
\newblock ``Sensei: sensing educational interaction,''
\newblock {\em UbiComp}, vol. 1, no. 4, pp. 1--27, 2018.

\bibitem{ahuja2019edusense}
Karan Ahuja, Dohyun Kim, Franceska Xhakaj, Virag Varga, Anne Xie, Stanley
  Zhang, Jay~Eric Townsend, Chris Harrison, Amy Ogan, and Yuvraj Agarwal,
\newblock ``Edusense: Practical classroom sensing at scale,''
\newblock {\em UbiComp}, vol. 3, no. 3, pp. 1--26, 2019.

\bibitem{littlewort2011computer}
Gwen Littlewort, Jacob Whitehill, Tingfan Wu, Ian Fasel, Mark Frank, Javier
  Movellan, and Marian Bartlett,
\newblock ``The computer expression recognition toolbox (cert),''
\newblock in {\em FG}. IEEE, 2011, pp. 298--305.

\bibitem{zeng2020emotioncues}
Haipeng Zeng, Xinhuan Shu, Yanbang Wang, Yong Wang, Liguo Zhang, Ting-Chuen
  Pong, and Huamin Qu,
\newblock ``Emotioncues: Emotion-oriented visual summarization of classroom
  videos,''
\newblock {\em TVCG}, vol. 27, no. 7, pp. 3168--3181, 2020.

\bibitem{ahuja2021classroom}
Karan Ahuja, Deval Shah, Sujeath Pareddy, Franceska Xhakaj, Amy Ogan, Yuvraj
  Agarwal, and Chris Harrison,
\newblock ``Classroom digital twins with instrumentation-free gaze tracking,''
\newblock in {\em CHI}, 2021, pp. 1--9.

\bibitem{picard2011measuring}
Rosalind~W Picard,
\newblock ``Measuring affect in the wild,''
\newblock in {\em ACII}. Springer, 2011, pp. 3--3.

\bibitem{hassib2017engagemeter}
Mariam Hassib, Stefan Schneegass, Philipp Eiglsperger, Niels Henze, Albrecht
  Schmidt, and Florian Alt,
\newblock ``Engagemeter: A system for implicit audience engagement sensing
  using electroencephalography,''
\newblock in {\em CHI}, 2017, pp. 5114--5119.

\bibitem{lin2018hand}
Jiaojiao Lin, Fei Jiang, and Ruimin Shen,
\newblock ``Hand-raising gesture detection in real classroom,''
\newblock in {\em ICASSP}. IEEE, 2018, pp. 6453--6457.

\bibitem{zheng2020intelligent}
Rui Zheng, Fei Jiang, and Ruimin Shen,
\newblock ``Intelligent student behavior analysis system for real classrooms,''
\newblock in {\em ICASSP}. IEEE, 2020, pp. 9244--9248.

\bibitem{ramakrishnan2021toward}
Anand Ramakrishnan, Brian Zylich, Erin Ottmar, Jennifer LoCasale-Crouch, and
  Jacob Whitehill,
\newblock ``Toward automated classroom observation: Multimodal machine learning
  to estimate class positive climate and negative climate,''
\newblock {\em TAC}, 2021.

\bibitem{bradski2008learning}
Gary Bradski and Adrian Kaehler,
\newblock {\em Learning OpenCV: Computer vision with the OpenCV library},
\newblock " O'Reilly Media, Inc.", 2008.

\bibitem{tomar2006converting}
Suramya Tomar,
\newblock ``Converting video formats with ffmpeg,''
\newblock {\em Linux Journal}, vol. 2006, no. 146, pp. 10, 2006.

\bibitem{russell2008labelme}
Bryan~C Russell, Antonio Torralba, Kevin~P Murphy, and William~T Freeman,
\newblock ``Labelme: a database and web-based tool for image annotation,''
\newblock {\em IJCV}, vol. 77, no. 1-3, pp. 157--173, 2008.

\bibitem{everingham2010pascal}
Mark Everingham, Luc Van~Gool, Christopher~KI Williams, John Winn, and Andrew
  Zisserman,
\newblock ``The pascal visual object classes (voc) challenge,''
\newblock {\em IJCV}, vol. 88, no. 2, pp. 303--338, 2010.

\bibitem{jocher2020yolov5}
Glenn Jocher, K~Nishimura, T~Mineeva, and R~Vilari{\~n}o,
\newblock ``Yolov5,''
\newblock {\em Code repository https://github.com/ultralytics/yolov5}, 2020.

\bibitem{ren2015faster}
Shaoqing Ren, Kaiming He, Ross Girshick, and Jian Sun,
\newblock ``Faster r-cnn: Towards real-time object detection with region
  proposal networks,''
\newblock {\em NIPS}, 2015.

\bibitem{dai2016r}
Jifeng Dai, Yi~Li, Kaiming He, and Jian Sun,
\newblock ``R-fcn: Object detection via region-based fully convolutional
  networks,''
\newblock {\em NIPS}, 2016.

\bibitem{zhang2016joint}
Kaipeng Zhang, Zhanpeng Zhang, Zhifeng Li, and Yu~Qiao,
\newblock ``Joint face detection and alignment using multitask cascaded
  convolutional networks,''
\newblock {\em SPL}, vol. 23, no. 10, pp. 1499--1503, 2016.

\bibitem{acharya2018covariance}
Dinesh Acharya, Zhiwu Huang, Danda Pani~Paudel, and Luc Van~Gool,
\newblock ``Covariance pooling for facial expression recognition,''
\newblock in {\em CVPRW}, 2018, pp. 367--374.

\bibitem{zhou2018raising}
Huayi Zhou, Fei Jiang, and Ruimin Shen,
\newblock ``Who are raising their hands? hand-raiser seeking based on object
  detection and pose estimation,''
\newblock in {\em ACML}. PMLR, 2018, pp. 470--485.

\bibitem{narasimhaswamy2022whose}
Supreeth Narasimhaswamy, Thanh Nguyen, Mingzhen Huang, and Minh Hoai,
\newblock ``Whose hands are these? hand detection and hand-body association in
  the wild,''
\newblock in {\em CVPR}, 2022, pp. 4889--4899.

\bibitem{kreiss2021openpifpaf}
Sven Kreiss, Lorenzo Bertoni, and Alexandre Alahi,
\newblock ``Openpifpaf: Composite fields for semantic keypoint detection and
  spatio-temporal association,''
\newblock {\em TPAMI}, 2021.

\bibitem{lin2014microsoft}
Tsung-Yi Lin, Michael Maire, Serge Belongie, James Hays, Pietro Perona, Deva
  Ramanan, Piotr Doll{\'a}r, and C~Lawrence Zitnick,
\newblock ``Microsoft coco: Common objects in context,''
\newblock in {\em ECCV}. Springer, 2014, pp. 740--755.

\end{thebibliography}

\end{document}